\documentclass[aps,showpacs,twocolumn,superscriptaddress,longbibliography]{revtex4-2}

\usepackage{lipsum}
\usepackage{physics}
\usepackage{amsmath}
\usepackage{amsfonts}
\usepackage{amssymb}
\usepackage{epsfig}
\usepackage{bm}

\usepackage{float}
\usepackage{braket}

\usepackage{graphicx}

\usepackage{bbold}

\usepackage{hyphenat}
\usepackage[colorlinks = true,
            linkcolor = blue,
            urlcolor  = blue,
            citecolor = blue,
            anchorcolor = blue]{hyperref}

\usepackage{placeins}
\usepackage{xcolor}

\usepackage{array}
\newcolumntype{C}[1]{>{\centering\arraybackslash}m{#1}}
\usepackage{overpic}

\usepackage{soul}

\renewcommand{\vec}{\mathbf}

\begin{document}

\title{Quantum geometry and low-frequency optical conductivity of nodal planes}

\date{\today}

\author{Raymond Wiedmann}
\email{r.wiedmann@fkf.mpg.de}
\affiliation{Max-Planck-Institut für Festkörperforschung, Heisenbergstraße 1, 70569 Stuttgart, Germany}

\author{Kirill Alpin}
\affiliation{Max-Planck-Institut für Festkörperforschung, Heisenbergstraße 1, 70569 Stuttgart, Germany}

\author{Moritz M. Hirschmann}
\affiliation{RIKEN Center for Emergent Matter Science, Wako, Saitama, 351-0198, Japan}

\author{Andreas P. Schnyder}
\affiliation{Max-Planck-Institut für Festkörperforschung, Heisenbergstraße 1, 70569 Stuttgart, Germany}


\begin{abstract}
        Nodal planes, two-dimensional symmetry-enforced band crossings, can carry a topological charge, similar to Weyl points. While the transport properties of Weyl points are well understood, those of nodal planes remain largely unexplored. These properties are influenced not only by the Berry curvature, but also by other quantum geometric quantities. In this work we study the quantum geometry — specifically the Berry curvature and quantum metric — and the linear optical conductivity of topological nodal planes. We introduce a low-energy model and investigate its low-frequency optical responses to determine
        the unique signatures of nodal planes.
        By comparing these findings to the optical response in a tight-binding model with a topological nodal plane, we observe consistent low-frequency behavior with a cubic power law.  This paves the way for the experimental detection of nodal planes through optical conductivity measurements for which we suggest suitable materials, most promisingly the material group $X$Mo$_3$S$_3$ ($X$ = Rb, K, Cs).
\end{abstract}

\maketitle

\section{Introduction}

Topological semimetals represent a fascinating class of quantum materials characterized by their unique electronic structures, where conduction and valence bands touch at discrete points, along lines, or on planes in momentum space \cite{Yan_2017,Bernevig_2018_review}.
Unlike conventional semimetals, topological semimetals exhibit robust surface states and unusual bulk properties that are protected by topological invariants. These materials, including Weyl and Dirac semimetals, have garnered significant interest due to their potential applications in electronics \cite{Grushin2017_Viewpoint}, spintronics \cite{parkin_spintronics_IEEE_2003,Kohno2021}, and quantum computing \cite{kharzeev2019chiralqubitquantumcomputing}.
For example, ideas on employing the chiral anomaly and the Weyl crossings in Weyl semimetals for chiral electronic devices have been proposed \cite{Kharzeev2013}.

The experimental discovery and characterization of topological semimetals often involves a combination of theoretical predictions and sophisticated measurement techniques \cite{RevModPhys.93.025002}. Angle-resolved photoemission spectroscopy (ARPES) \cite{Liu_2014, Lv_2015} and quantum oscillation measurements~\cite{wilde_Nature_21,huber_alpin_CoSi_PRL_22,huber_CoSi_PRB_24} are commonly used to probe their electronic structure. In addition, optical response measurements offer a complementary approach and have been used to find evidence of Dirac and Weyl points \cite{Pronin_2020}. 
These point crossings give rise to
characteristic frequency dependences
in the optical conductivity
\cite{Xu2016_WeylOpt, KunzeKoepfCaoetal.2024, Rodriguez_2020} and distinctive (quantized) features in photogalvanic responses~\cite{deJuan2016QuantizedCP,Rees2020,Ni_2021, Elio2017}.

Optical response measurements are particularly useful because they can be performed over a wide range of frequencies, temperatures, and external conditions,
and do not require big samples or large-scale measurement facilities.
The versatility of optical measurements allows for the exploration of various physical phenomena and a  fast screening of a large number of samples,  potentially enabling the discovery of new topological systems. Therefore, a better understanding of the optical responses does not only deepen our fundamental knowledge of topological semimetals, but
is also useful to screen material candidates and for the development of novel applications based on topological  responses.

While the optical responses of Weyl and Dirac points are relatively well understood, those of nodal planes remain mostly unknown. This is in part because topological semimetals with nodal planes, i.e., degeneracies on two-dimensional (2D) planes in the Brillouin zone (BZ), have only recently gained attention \cite{zhong2016towards, Liang2016, bzduvsek2017robust, Wu2018,T_rker_2018, zhang2018nodal, yu2019circumventing, fu2019dirac, chang2018topological, Salerno_2020_floquet, wilde_Nature_21, huber_alpin_CoSi_PRL_22, ma2021observation, tang2022complete, alpin2023fundamental, frank2024weyl, chen2022large, khanh2024gappednodalplanesdrive, li2025giant, yamada2025gapping, hu2023hierarchy, xie2021three, Xiao2020chargednodalsurface, yang2019observation, kim2019topologically, scheie2022dirac, agterberg2017bogoliubov, ohashi2024surface}.
The advantage compared to topological nodal features of lower dimension, like Weyl points, is that nodal planes cover a larger part of the BZ and, if dispersive, a large energy range in the spectrum. This means that less fine tuning is needed in order for them to be part of the Fermi surface. Recently, a large Nernst response of (weakly gapped) topological nodal planes has been detected \cite{khanh2024gappednodalplanesdrive}. This raises the question whether nodal plane materials generically show particularly large or characteristic optical responses, which  can be traced back to their nontrivial topology and characteristic quantum geometry.

In this work, we provide a simple, but generic, low-energy continuum model of a topological nodal plane which can be used to tackle this question. We first discuss the properties of the band structure of the model before analyzing its quantum geometry, i.e., the Berry curvature and quantum metric. The linear optical conductivity is directly determined by the quantum metric. We analytically calculate its component perpendicular to the nodal plane and find a generic $\omega ^3$ power law for low frequencies $\omega$.
To verify the generality of this low-energy power law, we then numerically compute the optical conductivity of a generic tight-binding model with a nodal plane and find
very good agreement with the analytics.
Finally, we discuss requirements for the experimental detection of the signature of nodal planes in optical conductivity measurements and list suitable materials.


\section{Low-energy model of a topological nodal plane}
A stable nodal degeneracy on a 2D surface, called nodal plane, in the BZ can be enforced by the combined symmetry $\Tilde{\mathcal{C}}_2^{a}\mathcal{T}$. It consists of a two-fold screw rotation $\Tilde{\mathcal{C}}_2^{a}$ around the axis $a$ with a non-symmorphic lattice translation by $\frac{1}{2}$ of the unit cell along the rotation axis and time-reversal symmetry $\mathcal{T}$ \cite{Liang2016, Xiao2020chargednodalsurface}. One finds that $\Tilde{\mathcal{C}}_2^{a}\mathcal{T}$ squares to $-1$ at $k_a = \pi$ leading to a two-fold degeneracy for all points on the $k_a = \pi$ plane due to Kramers theorem.

\begin{figure*}
    \centering
    \includegraphics[width=\linewidth]{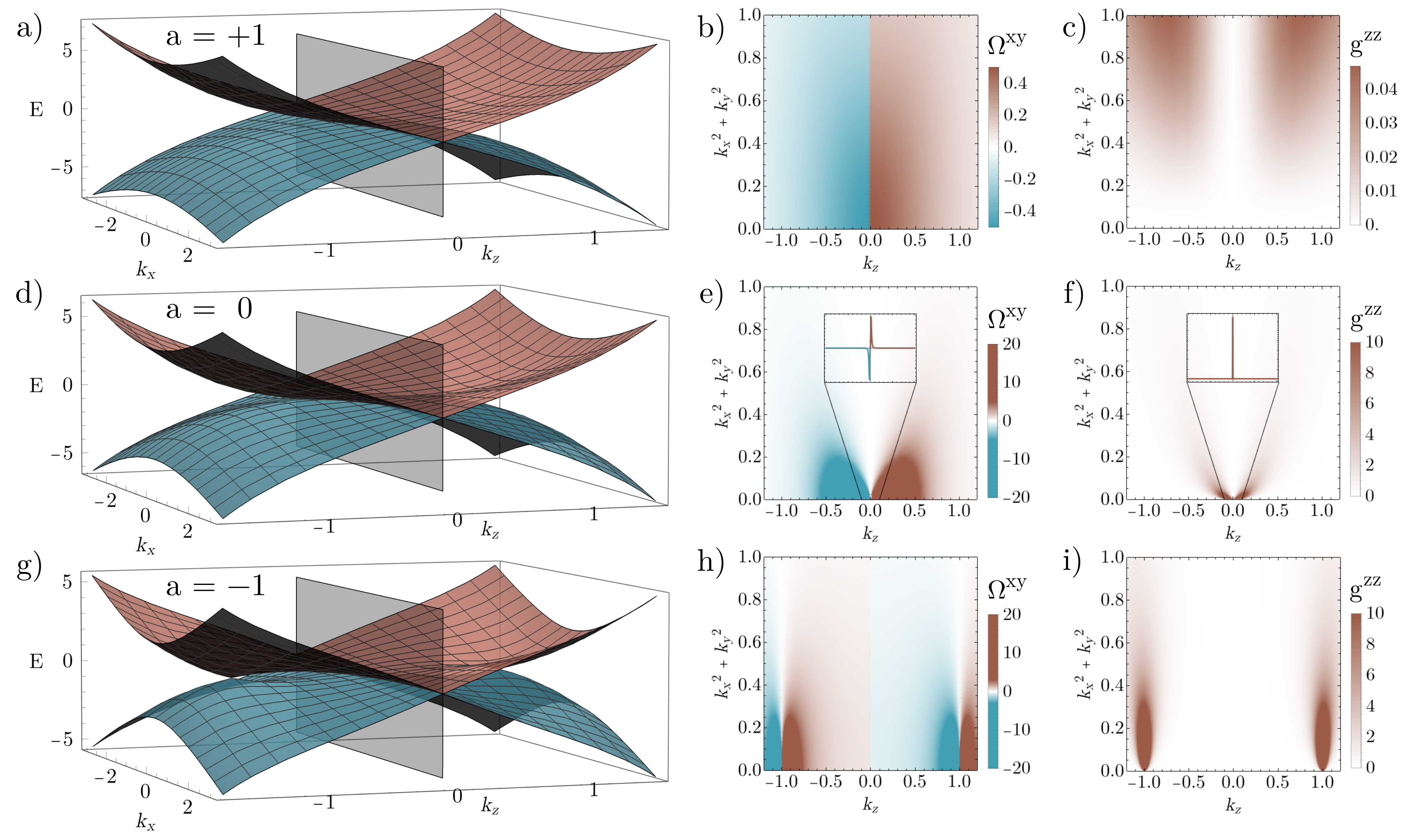}
    \caption{
    (a), (d), (g)
    Band structure of the low-energy model, Eq.~\eqref{Eq:LE_NP}, at $k_y=0$ with $\varepsilon=1$ for $a=1$, $a=0$, and $a=-1$, respectively. The position of the nodal plane is marked by a gray plane indicating its extension in the $k_y$-direction. The distribution of Berry curvature $\Omega^{xy}$ in the model for the three cases is shown in (b), (e), (h) and the distribution of quantum metric $g^{zz}$ in (c), (f), (i). Note that the distribution of quantum geometry only depends on $k_z$ and the radial distance $k_x^2 + k_y^2$ which is used in these figures. For the figures showing the distribution of quantum geometry, the scales were cut in order for the details to be visible. The fact that these quantities are diverging in the cases of $a = 0$ (and $a = -1$) is indicated in the insets of (e) and (f) (and hold also for (h) and (i)).
    }
    \label{fig:LE_bands_geometry}
\end{figure*}

For the construction of a low-energy model of a topological nodal plane, the only restriction is that it must be symmetric under this $\Tilde{\mathcal{C}}_2^a \mathcal{T}$ symmetry, where we choose $a=z$ for the rotation axis,
leading to a nodal plane at $k_z = 0$. The $\Tilde{\mathcal{C}}_2^{z}\mathcal{T}$ symmetry acts on the Pauli matrices as $\sigma_i \rightarrow - \sigma_i$ for $i \in \{x,y,z\}$ and on the momenta as $k_{x/y} \rightarrow  k_{x/y}$ and $k_{z} \rightarrow  - k_{z}$.
Therefore, for the symmetry to be fulfilled, every term in the Hamiltonian, except for trivial terms $\sim \mathbb{1}$, has to be odd in $k_z$. Up to cubic order in $k$, we obtain the Hamiltonian
\begin{align}\label{Eq:LE_NP}
    \mathcal{H}(\mathbf{k}) = \varepsilon k_x k_z \sigma_x + \varepsilon k_y k_z \sigma_y + \varepsilon ( k_z^3 + a k_z) \sigma_z \; ,
\end{align}
which fulfills these requirements and exhibits a flat topological nodal plane at $k_z=0$ with charge $\nu = \text{sign}(\varepsilon a)$. The Pauli matrices are denoted by $\sigma_i$ and the parameter $\varepsilon$ determines the dispersion of the bands away from the nodal plane,
i.e., in the direction perpendicular to the nodal plane.
In particular, close to the nodal plane and for $a\neq 0$ the dispersion is linear with the slope $\varepsilon a$. We set $\varepsilon = 1$ for this discussion. Since the charge of the nodal plane depends on $a$, by varying it, one can observe a phase transition between $\nu = +1$ and $\nu = -1$, which is accompanied by the creation of two Weyl points. The case $a=0$ exhibits a special behaviour, which is discussed in more detail in the following sections. In terms of the band structure, however, it is qualitatively the same as for $a>0$.  The band structure for the three cases, $a=+1$, $a=0$, and $a=-1$, are shown in Figs.~\ref{fig:LE_bands_geometry} (a),(d),(g), respectively. Note that this low-energy model is not unique, since not all allowed cubic terms are included. It is, however, the simplest model which exhibits the generic features discussed in the following sections. An argument for this and the discussion of low-energy models of different order in $k$
are given in Appendix~\ref{app:lowenergymodels}.

\section{Quantum geometry of the low-energy model}
In recent years, the quantum geometry of wave functions has received increasing attention, which has revealed important
connections between optical response functions and
quantum geometric quantities
\cite{Ahn2021_Riemannian, avdoshkin2024multistategeometryshiftcurrent,Ezawa_2024,ghosh2024probingquantumgeometryoptical,jiang2025revealingquantumgeometrynonlinear,slagerPRB2025,WeiChenSumRule2024,WeiChenOpticaMarker2025}. In the following, we will give a very brief review on the topic of quantum geometry (for a perspective on the field see \cite{T_rm__2023}, for a review \cite{Liu_2024_review}) before discussing the relevant geometric features of the nodal plane low-energy model, i.e., the Berry curvature and the quantum metric. We will restrict our discussion to two-band models, which is sufficient for the study of the systems in this paper.

Generically, through diagonalization of a Hamiltonian, one obtains eigenvalues, the energy bands, and the corresponding eigenstates, the wave functions. In many cases, in order to understand the basic behaviour of materials, such as whether they are insulating or metallic, the study of the eigenvalues is sufficient. However, with the rise of topological systems, the significance of wave function properties reflected in the Berry curvature, i.e., the phase distance between quantum states, has become clear. The Berry curvature is, however, only one aspect of the geometry of the wave function. Recently, a lot of attention has focused on the quantum metric, which quantifies the change in the amplitude of the wave function \cite{MitscherlingHolder2022, Ozawa_2021,Wang_2023, Huhtinen_2022, li2024flatbandjosephsonjunctions,kruchkov2024entanglemententropylatticemodels,MitscherlingMera2022,WeiChenMetric2021}. The Berry curvature and quantum metric are contained in the quantum geometric tensor (QGT) \cite{cheng2013quantumgeometrictensorfubinistudy}.

To discuss the QGT we consider a generic two-band model described by the Hamiltonian
\begin{align}
    H(\vb{k})
    =
    d_0(\vb{k}) \mathbb{1} + \vb{d}(\vb{k}) \cdot \bm{\sigma} ,
\end{align}
with $\vb{d}(\vb{k}) = (d_x,d_y,d_z)^T, d_i \in \mathbb{R}$ ($i \in \{0, x,y,z\}$) and the vector of Pauli matrices $ \bm{\sigma}$. The energies of the two bands are $E_\pm = d_0(\vb{k}) \pm \vert\vb{d}(\vb{k})\vert$. One can define a normalized vector $\vec n (\vec k) = \frac{\vb{d}(\vb{k})}{\vert\vb{d}(\vb{k})\vert}$ with which the QGT of the two bands denoted by $\pm$ can be expressed as
\begin{align}\label{Eq:QGT}
    \mathcal{Q}^{ab}_\pm &= \frac{1}{4} \partial_a \vec n \cdot \partial_b \vec n \pm \frac{i}{4} \vec n \cdot (\partial_a \vec n \times \partial_b \vec n)  \\
    &= g^{ab}_\pm - \frac{i}{2} \Omega^{ab}_\pm \nonumber \; ,
\end{align}
where the real part corresponds to the quantum metric $g^{ab}_\pm$ and the imaginary part to the Berry curvature $\Omega^{ab}_\pm$. Note that we have dropped the $\vec k$ dependence for simplicity and the derivatives are denoted with respect to the $k_a$ direction, i.e., $\partial_a \equiv \partial_{k_a}$. The QGT is the lowest-order quantum geometric quantity.
Higher-order objects have been described theoretically and are connected, for example, to higher-order optical responses \cite{Ahn2021_Riemannian, avdoshkin2024multistategeometryshiftcurrent}. Here we are interested in the linear conductivity, for which only the QGT is required.

With this, we can discuss the quantum geometry of the nodal plane low-energy model defined in Eq.~\eqref{Eq:LE_NP} for the three cases of $a$ discussed above. However, for a better intuition, we will briefly review the case of a Weyl point described by the low-energy model $\mathcal{H}_\mathrm{WP} = \vec k \cdot \bm{\sigma}$. For this 0D nodal feature it is well known that the Berry curvature given by $\vec \Omega_\pm = \mp \frac{\vec k}{2 (k_x^2 + k_y^2 + k_z^2)^\frac{3}{2}}$ diverges at the crossing and the charge of the Weyl point can be calculated by determining the flux of Berry curvature through a sphere enclosing it. In fact, also the quantum metric diverges at that point.

The nodal plane cannot be enclosed by a sphere since it extends infinitely in the $k_x$-$k_y$ plane. In order to obtain its charge, we calculate the flux of Berry curvature through two planes at $k_z = \pm \xi$, which only encompass the nodal plane, i.e., we consider the limit $\xi \rightarrow 0$. The relevant Berry curvature component is the one perpendicular to the nodal plane, here $\Omega^{xy}$, which has a sign jump at the nodal plane. This yields
\begin{align}
    C(k_x,k_y) &= \lim_{\xi\rightarrow0} \Omega^{xy} (k_x,k_y,\xi) - \Omega^{xy} (k_x,k_y,-\xi) \nonumber \\
    &= \frac{\mathrm{sign}(\varepsilon)\  a }{(a^2+k_x^2+k_y^2)^\frac{3}{2}} ,
\end{align}
and the resulting charge is $\nu = \frac{1}{2\pi}\int \mathrm{d}k_x \mathrm{d}k_y C(k_x,k_y)$. $C(k_x,k_y)$ can be interpreted as a topological charge density on the nodal plane. Here, it also becomes obvious that the parameter $a$ determines the spread of Berry curvature over the nodal plane. The total charge, however, only depends on the sign of $a$ and of $\varepsilon$. Apart from this, the parameter $\varepsilon$ has no impact on the geometric quantities.

In Figs.~\ref{fig:LE_bands_geometry} (b),(e),(h) we show the distribution of the Berry curvature of the entire system for the three choices of $a$ discussed in the previous section. One finds that for $a=1$ the Berry curvature has the expected sign jump at the nodal plane and is spread broadly in the $k_x^2 + k_y^2$-plane. The charge is $\nu = +1$. The case $a=0$ is a special point since here the Berry curvature diverges at $\vec k = 0$, as indicated in the inset of Fig.~\ref{fig:LE_bands_geometry}(e), which resembles the monopole behaviour of a Weyl point and, as for $a=1$, it integrates to $\nu = +1$. For $a=-1$ two additional Weyl points display the characteristic divergence of Berry curvature and the sign jump at the nodal plane is reversed, leading to a charge of $\nu = -1$. We find that the $\Omega^{xy}$ component is sufficient to study the topological properties of the nodal plane, since its behaviour is determined by the symmetry enforcing the nodal plane.

The quantum metric, as shown in Eq.~\eqref{Eq:QGT}, is determined by the derivatives of the normalized vector $\vec n (\vec k)$. As for the Berry curvature, we therefore expect the component perpendicular to the nodal plane $g^{zz}$ to be the most generic. However, since there is no quantization of the integral of individual quantum metric components, usually only qualitative statements about its behaviour can be made. The $zz$-component is given by
\begin{align}
    g^{zz}_\pm = \frac{(k_x^2 + k_y^2) k_z^2}{(k_x^2 + k_y^2 + (a+k_z^2)^2)^2},
\end{align}
which is positive definite and symmetric in $k_z$. The distribution of $g^{zz}$ is, as for the Berry curvature, shown for the three cases of $a$ in Figs.~\ref{fig:LE_bands_geometry} (c),(f),(i).
We find again that its distribution depends on $a$. In particular, for $a=1$, the metric is broadly distributed around the nodal plane. For $a=0$ the metric collapses to $\vec k = 0$, as indicated in the inset of Fig.~\ref{fig:LE_bands_geometry}(f), showing the discussed Weyl point behaviour. For $a=-1$ the two emerging Weyl points again dominate the metric distribution. This shows that the distribution roughly follows the one of the Berry curvature. Therefore, topological nodal features, such as Weyl points and nodal planes, can be seen as the source of non-trivial quantum geometry.
Importantly, this non-trivial behaviour is not   restricted to signatures
of the Berry curvature or quantum metric. Rather, we expect that, depending on the symmetries of the system,  this will also show up in higher-order quantum geometric quantities, such as the quantum connection \cite{Ahn2021_Riemannian,mitscherling2024gaugeinvariantprojectorcalculusquantum,avdoshkin2024multistategeometryshiftcurrent,jiang2025revealingquantumgeometrynonlinear,slagerPRL2024}. The discussion of this is, however, beyond the scope of this work.

\section{Linear optical conductivity}
\label{sec:linoptcond}
Knowing the quantum geometric properties of the nodal plane low-energy model, we can now study the effects of this non-trivial behaviour in responses. In particular, we are interested in the signatures of this in the optical conductivity.
By expanding the current density for small electric fields $E_i$ one obtains \cite{Ahn2020_Lowfreqdiv,Ahn2021_Riemannian}
\begin{align}
    j^c = \sigma_{(1)}^{ca} E_a + \sigma_{(2)}^{cab} E_a E_b + ... \; ,
\end{align}
where the first term is the linear response and all higher terms give non-linear responses. Here, we are interested in the linear optical response with the optical conductivity tensor $ \sigma_{(1)}^{ca}$.

For a two-band model, the interband linear conductivity can be expressed in terms of the QGT as \cite{Ahn2021_Riemannian}
\begin{align}
    \sigma^{ab} = \frac{\pi \omega e^2}{\hbar} \int \frac{\text{d}^3 k}{(2 \pi)^3} \delta \left(\omega - \frac{\Delta E}{\hbar}\right) \Delta f \mathcal Q_{\pm}^{ab},
\end{align}
with the QGT $\mathcal Q_{\pm}^{ab}$ defined in Eq.~\eqref{Eq:QGT}, $\Delta f= f_- - f_+$ being the difference of the Fermi-Dirac distributions of the two bands, and $\Delta E = E_+ - E_-$ the energy gap between the two bands.
We are interested in the real part of the optical conductivity given by \cite{Bacsi2013}
\begin{align}\label{eq:linoptcond}
    \text{Re}[\sigma^{ab}] =  \frac{\pi \omega e^2}{4 \hbar (2 \pi)^3} \int \text{d}^3 k \delta \left(\omega - \frac{\Delta E}{\hbar}\right)  \Delta f\, \partial_a \Vec{n} \cdot \partial_b \Vec{n} ,
\end{align}
which is directly determined by the quantum metric. We set $\hbar=e=1$ in the following.

\begin{figure}
    \centering
    \includegraphics[width=\linewidth]{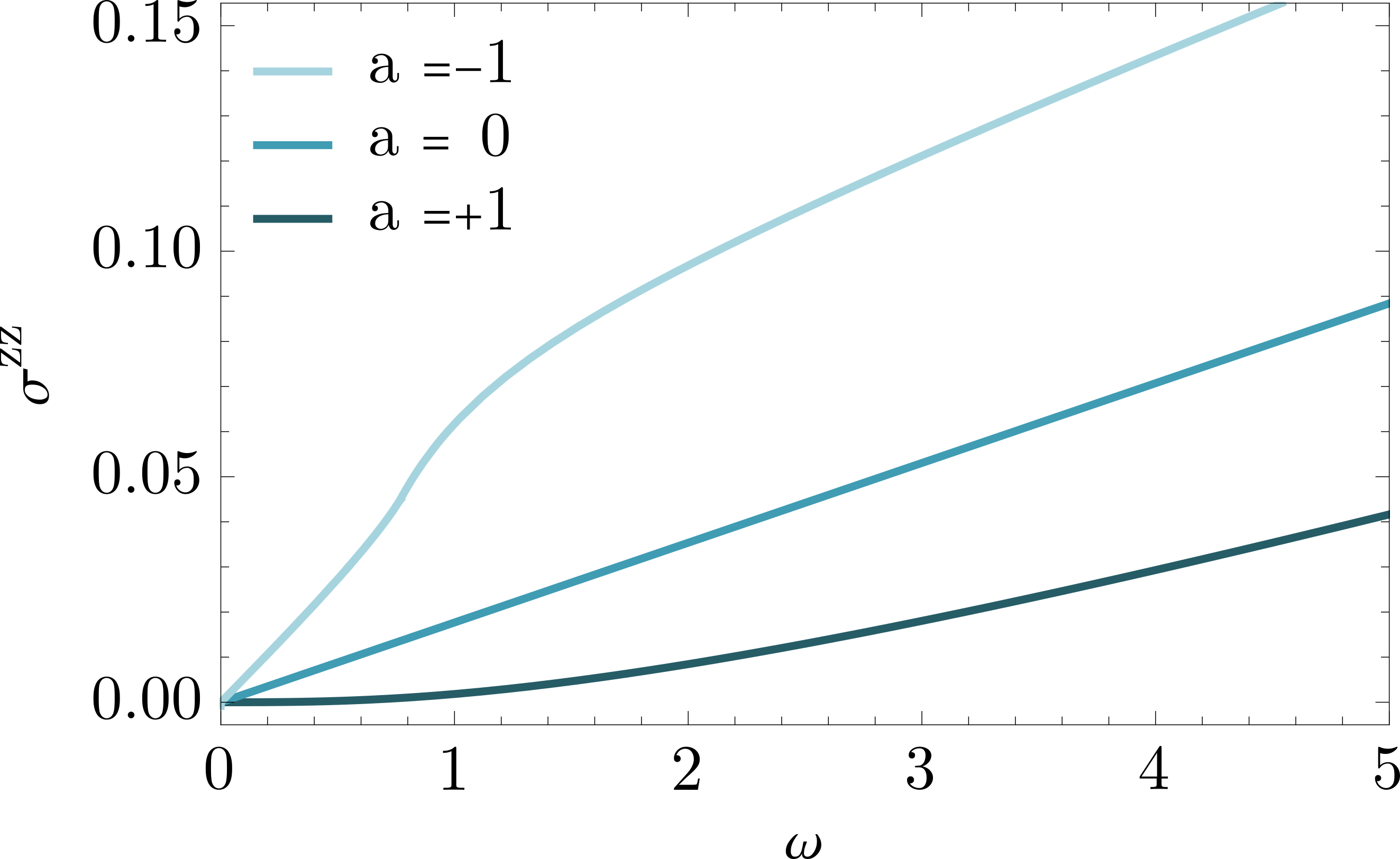}
    \caption{
    $zz$-component of the interband linear conductivity, $\sigma^{zz}$,
    for the low-energy model, Eq.~\eqref{Eq:LE_NP},
    as a function of frequency $\omega$
    for three different values of $a$ with $\varepsilon=1$.
    }
    \label{fig:linearcond_zz}
\end{figure}

As discussed for the quantum geometry, we will focus on the optical conductivity component perpendicular to the nodal plane $\sigma^{zz}$, which is determined by the quantum metric component $g^{zz}$. Only this component is restricted by the $\Tilde{\mathcal{C}}_2^{z}\mathcal{T}$ symmetry, and therefore expected to show generic features. To evaluate this for the nodal plane low-energy model at $T=0$, we consider the cases $a\geq 0$ and $a<0$ separately. In Fig.~\ref{fig:linearcond_zz}, the results for the different values of parameter $a$ are shown. The full analytical expression for $a\geq 0$ is given in Appendix~\ref{app:analytcond} and, in the following, the leading-order behaviour is discussed. We find that for $a>0$ the leading order in $\omega$ in the Taylor expansion of the conductivity is cubic, i.e.,
\begin{align}
    \sigma^{zz}_{a>0} = \frac{ \pi^2}{15 (2 \pi)^3(a\varepsilon)^3} \omega^3 - \frac{ \pi^2}{28 (2 \pi)^3 a^6 \varepsilon^5}\omega^5 +  \mathcal{O}(\omega^7)\; ,
\end{align}
and all even orders in $\omega$ vanish. The limiting case of $a=0$ shows an exact linear dependence
 \begin{align}
     \sigma^{zz}_{a=0} = \frac{ 1}{ (2 \pi)^3} \frac{4 \pi^2}{9 |\varepsilon|} \omega\; ,
 \end{align}
 aligning with the analysis of the quantum geometry in that case, which suggests a Weyl point behaviour. It is well known that Dirac and Weyl points show such a linear-in-frequency dependence \cite{Hosur_2012}.
In the case of $a<0$, where two additional Weyl points arise, we determine the optical conductivity numerically. The result  shown in Fig.~\ref{fig:linearcond_zz} indicates a linear leading-order behaviour for small $\omega$, which is in line with the analysis of the quantum geometry for $a<0$, where the Weyl points are dominating the Berry curvature and quantum metric distribution.
 At higher frequencies, the contributions from the nodal plane lead to a stronger deviation from the linear behaviour. Interestingly, the limit $\omega\rightarrow\infty$ always yields the same result, which is exactly $\sigma^{zz}_{a=0}$. This can be understood such that if one moves `far away' from the nodal plane, the exact distribution of quantum metric will not matter anymore, and it will behave as a Weyl point, meaning that the geometry behaves as a monopole. However, in this limit, our low-energy model is not valid anymore.

\begin{figure*}
    \centering
    \includegraphics[width=\linewidth]{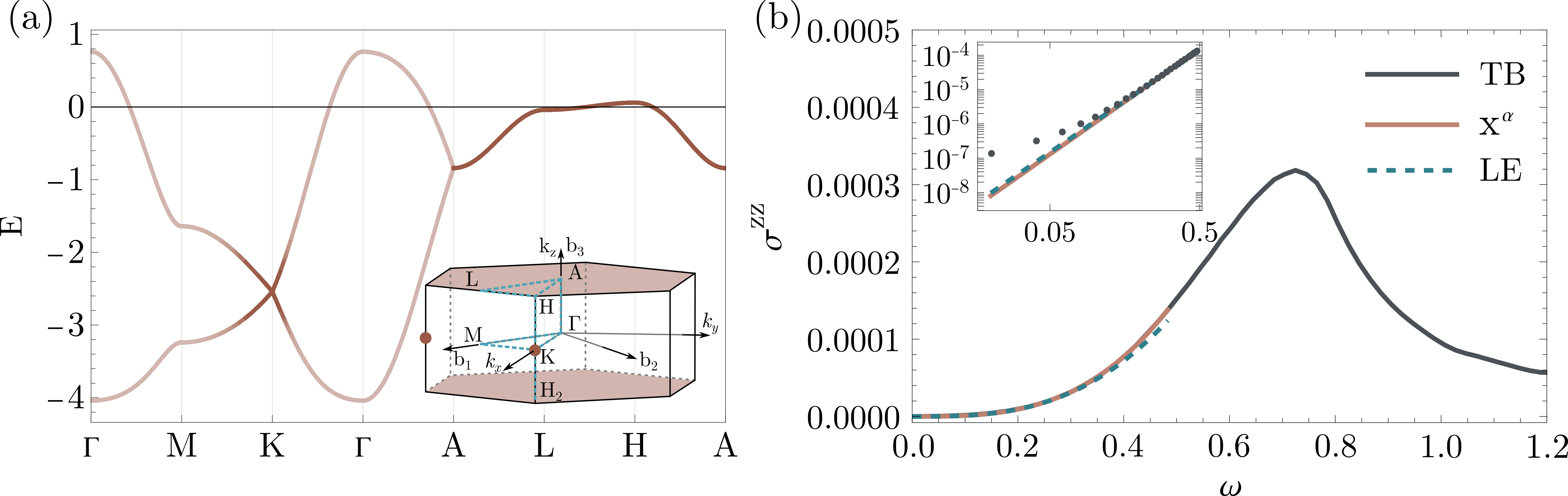}
    \caption{(a) Band structure of the modified tight-binding model introduced in \cite{Xiao2020chargednodalsurface} with a topological nodal plane at $k_z = \pi$ and a two Weyl points at the K points.  The opacity of the bands indicates the areas of large (diverging) $zz$-component of the quantum metric. The inset shows the hexagonal BZ with the high-symmetry points. (b) $zz$-component of the linear optical conductivity for the tight-binding (TB) model calculated numerically using a phenomenological broadening of $\Gamma = 0.02$. A fit of the form $x^\alpha$ in the low-frequency region ($\omega \leq 0.5$) yields the power-law behaviour $\alpha = 3.07 \pm 0.01$ and is indicated with a pink line. The analytical result for the low-energy model with the parameters of the tight-binding model is shown for low frequencies with the dashed line (LE). The inset shows the low-frequency behaviour using logarithmic scales on both axes.
    }
    \label{fig:tblinearcond_zz}
\end{figure*}

While the results for $a\leq0$ are interesting and in agreement with an intuitive understanding of these topological features, the $\omega^3$ dependence for $a>0$, i.e., the case with a single topological nodal plane, is completely novel and, due to the enforcing symmetry that impacts the $zz$-component of the quantum metric, expected to be a characteristic signature of a topological nodal plane.
Higher-order terms in $k$ in the low-energy description can only impact higher-order terms in the $\omega$-expansion. This means that, if a material has a single topological nodal plane at the Fermi energy (without additional topological features nearby), this characteristic frequency dependence can be used to detect the nodal plane, similarly to the detection of Weyl points through their linear behaviour \cite{Xu2016_WeylOpt}.

\section{Trivial nodal planes and arbitrary dispersions}
The characteristic cubic frequency dependence is not restricted to topologically charged nodal planes. A minimal low-energy model of a trivial nodal plane is given by
\begin{align}
    \mathcal{H}_\mathrm{triv} = k_z \sigma_x + k_z \sigma_y + (k_z + k_z^3)\sigma_z \; .
\end{align}
The only non-zero component of the QGT in the model is the quantum metric component perpendicular to the nodal plane given by
\begin{equation}
    g^{zz}_\pm = \frac{2 k_z^2}{(3+2k_z^2 + k_z^4)^2}\; .
\end{equation}
As for the topological nodal plane model, we calculate the linear optical conductivity, where in this case only the component perpendicular to the nodal plane $\sigma^{zz}$ is non-zero. The full expression is given in Appendix \ref{app:trivNP} and we obtain again a cubic power law in the low-frequency expansion given by $\sigma_{zz} = \frac{\pi}{(2\pi)^3 }\frac{\omega ^3}{54 \sqrt{3}} + \mathcal{O}\left(\omega ^5\right)$.

Furthermore, we show in Appendix \ref{app:dispersion} that our findings also hold when including arbitrary (symmetry-allowed) dispersion terms in the Hamiltonian, i.e., $\mathcal{H}_\text{disp} = \mathcal{H} + d_0(\mathbf{k}) \mathbb{1}$, since they enter the optical conductivity only through the Fermi-Dirac distribution. With this, also a variation of the chemical potential, corresponding to a constant $d_0$, cannot change this behaviour. This discussion holds on the level of low-energy models, while in tight-binding models other band features and the band width have to be taken into account. In particular, if other topological nodal features such as Weyl points 
occur in the vicinity of the Fermi energy,
these can dominate the low-frequency optical response.

\section{Optical conductivity of a tight-binding model}

In order to study, whether the characteristic low-frequency behaviour is visible also in more complicated systems, we consider a generic tight-binding model exhibiting a single nodal plane. Specifically, we examine a two-band tight-binding model for a hexagonal space group, which was introduced in \cite{Xiao2020chargednodalsurface},
see Appendix~\ref{app:tbmodel}.
We modify its dispersion by adding the term $-\cos(k_z)\mathbb{1}$ to increase the energy separation of the nodal plane and the Weyl points. This is crucial in order to study the signatures of the nodal plane alone, since, as shown in Sec. \ref{sec:linoptcond}, Weyl points close to the Fermi level will likely dominate the optical response. We choose the chemical potential such that a broad region of the nodal plane lies close to the Fermi energy. The resulting band structure is shown in Fig.~\ref{fig:tblinearcond_zz}(a). It has a nodal plane at $k_z = \pi$, which carries a topological charge $\nu=2$, and two Weyl points at the K points, which compensate the nodal-plane charge.

For the numerical evaluation of the optical conductivity of this model, a phenomenological broadening $\Gamma$ is introduced, i.e., the delta function $\delta (\omega - \frac{\Delta E}{\hbar})$ is replaced with a Lorentzian function $\frac{\Gamma}{\pi ((\omega - \frac{\Delta E}{\hbar})^2 + \Gamma^2)}$. The resulting interband linear optical conductivity $\sigma^{zz}$ is shown in Fig.~\ref{fig:tblinearcond_zz}(b) for a broadening of $\Gamma=0.02$. For low frequencies ($\omega \leq 0.5$) we show the exact analytical result obtained for the low-energy model using the parameters of the tight-binding model (for details see Appendix \ref{app:tbmodel}), and we perform a fit of the form $x^\alpha$ to extract the power-law behaviour, which yields $\alpha = 3.07 \pm 0.01$, agreeing with our expectations of a (larger than) cubic behaviour. The inset in Fig.~\ref{fig:tblinearcond_zz}(b) shows the low-frequency behaviour using logarithmic scales on both axes. We find that the power-law behaviour agrees very well except for very small frequencies comparable to the size of the broadening $\Gamma = 0.02$, where a deviation is expected. Further deviations of the numerical result of the tight-binding model and the analytical results for the low-energy model are caused by the non-uniform dispersion perpendicular to the nodal plane and the dispersion of the nodal plane itself in the tight-binding model. The low-energy model of the nodal plane is obtained by performing a series expansion around a single point of the nodal plane and can, therefore, not capture the exact quantitative behaviour.
 Qualitatively, however, we find that the low-frequency behaviour agrees very well.
Therefore, for this simple tight-binding model with a dispersive topological nodal plane, we can explain the low-frequency features with the results we obtain for a low-energy nodal plane. In particular, we find a $\omega^3$ behaviour in the component perpendicular to the nodal plane.

\section{Discussion and Material Candidates}
In this work we introduced a generic low-energy model for a topological nodal plane, which is enforced by $\Tilde{\mathcal{C}}_2^{z}\mathcal{T}$ symmetry. We found three distinct cases depending on the choice of parameter $a$, which exhibit a flat topological nodal plane and, in case of $a<0$, two additional Weyl points in the band structure. We then studied the quantum geometry, i.e., the Berry curvature and quantum metric, of this model, which shows the connection of a topological nodal plane to a Weyl point in the case of $a=0$. While the optical response of Weyl semimetals is well understood, the characteristic fingerprints of nodal planes in optical responses has not been studied so far. Here, we tackled this question by analytically determining the interband linear optical conductivity of the low-energy model, where we find a characteristic $\omega^3$ leading-order behaviour for $a>0$. We expect this behaviour to show up generically in the conductivity component perpendicular to the nodal plane. To support this claim, we considered a two-band tight-binding model with a single dispersive topological nodal plane at the Fermi level. The linear optical response of this system agrees well with the predicted $\omega^3$ behaviour for low frequencies.
Further, we found that this characteristic feature does not depend on a finite topological charge, but also shows up in a topologically trivial nodal plane low-energy model.

In order to understand, whether this signature can be seen in experiment, intraband contributions need to be considered as well. Depending on the size of the phenomenological broadening, the broadened Drude peak can overlap significantly with the low-frequency spectrum, making a reliable analysis of the behaviour challenging. Furthermore, we consider a model with a single nodal plane, which is particularly common in hexagonal systems \cite{Hirschmann_inPrep}. In general, the symmetries enforcing nodal planes can lead to nodal plane duos or trios~\cite{wilde_Nature_21}. In these cases, in-plane contributions to the optical conductivity from the different nodal planes are expected to lead also to lower-order contributions in $\omega$.

Additional topological nodal features, such as Weyl points, can also influence the low-frequency optical signature if they are in the vicinity the Fermi level. Therefore, systems with a single trivial nodal plane are expected to be better suited since the bands forming the nodal plane are not required to exhibit additional crossings to compensate the topological charge. In Appendix \ref{app:Materials}, we discuss the criteria for materials with which we expect the low-frequency signature to be visible most clearly and over a significant frequency range, we list all space groups with single nodal planes, and we provide suitable materials. In particular, we propose Sb$_3$Sr$_5$ \cite{Martinez73SbSr} and $X$Mo$_3$S$_3$ ($X$ = Rb, K, Cs) \cite{Wu2018,Huster1983333} to be promising materials for the measurement of the optical signature of nodal planes.

Our study adds simple low-energy models for topological and trivial nodal planes and predictions for their optical responses to the growing interest in the behaviour of nodal planes and how these could be harnessed in novel technologies. Future investigations of nonlinear optical responses and the effects of symmetry breaking will help further the understanding of these fascinating materials.

\begin{acknowledgments}
The authors thank Johannes Mitscherling for helpful discussions and insightful comments on the manuscript.
A.P.S., R.W., and K.A. are funded by the Deutsche Forschungsgemeinschaft (DFG, German Research Foundation) – TRR 360 – 492547816.
M.~M.~H.~is funded by the Deutsche Forschungsgemeinschaft (DFG, German Research Foundation) - project number 518238332.
\end{acknowledgments}

\section*{Data availability}
The data that support the findings of this article are openly available \cite{data_availability}.

\appendix
\renewcommand{\appendixname}{APPENDIX}

\section{Low-energy models of different order in $k$}
\label{app:lowenergymodels}

Nodal planes are enforced by $\Tilde{\mathcal{C}}_2^{a}\mathcal{T}$ symmetry, i.e., the combination of a two-fold screw-rotation $\Tilde{\mathcal{C}}_2^{a}$ around the axis $a$ with a translation by $\frac{1}{2}$ of the unit cell along the rotation axis and time-reversal symmetry $\mathcal{T}$. Here, we choose $z$ as the rotation axis.
This symmetry acts on the Pauli matrices as $\sigma_i \xrightarrow{\Tilde{\mathcal{C}}_2^{z}\mathcal{T}} - \sigma_i$ for $i \in \{x,y,z\}$ and on the momenta as $k_{x/y} \xrightarrow{\Tilde{\mathcal{C}}_2^{z}\mathcal{T}}  k_{x/y}$ and $k_{z} \xrightarrow{\Tilde{\mathcal{C}}_2^{z}\mathcal{T}}  - k_{z}$.
Therefore, for the symmetry to be fulfilled, every term in the Hamiltonian, except for trivial terms $\sim \mathbb{1}$, has to be odd in $k_z$.
The simplest possible low-energy model -- the one linear in $k$ -- which is $\Tilde{\mathcal{C}}_2^{z}\mathcal{T}$ symmetric and, therefore, exhibits a nodal plane is given by
\begin{align}
    \mathcal{H}^{(1)} = k_z \sigma_z \; .
\end{align}
This model is fully trivial in terms of its quantum geometry and all optical responses vanish. The inclusion of any higher-order symmetry-allowed terms will lead to a non-trivial model.

The lowest-order non-trivial model has been introduced in \cite{Xiao2020chargednodalsurface} and is given by
\begin{align}\label{eq:quadLEmodel}
    \mathcal{H}^{(2)} = k_z (k_x \sigma_x + k_y \sigma_x + a \sigma_z)\; ,
\end{align}
i.e., it is quadratic in $k$ and exhibits a topological nodal plane of charge $\nu = \mathrm{sign}(a)$. The transition from $\nu = +1$ to $\nu=-1$ is not accompanied by the generation of Weyl points, but a nodal line appears at $a=0$.
One finds that the quantum metric component orthogonal to the nodal plane $g_{zz}$ is zero everywhere. Therefore, also the corresponding optical conductivity component vanishes. Other components of the optical conductivity can be finite, however, due to the lack of symmetry restriction no generic behaviour is expected for them.

The model introduced in this paper
\begin{align}
    \mathcal{H}^{(3)} =  k_x k_z \sigma_x +  k_y k_z \sigma_y +  (k_z^3 + a k_z) \sigma_z \; ,
\end{align}
where we neglect additional prefactors for this discussion, is the simplest and lowest-order model to exhibit a finite quantum metric perpendicular to the nodal plane and, therefore, a finite $zz$-component of the linear optical conductivity.
Furthermore, its topological phase transition in accompanied by the creation of Weyl points, describing a more generic situation than the quadratic model in Eq.~\eqref{eq:quadLEmodel}.
Any additional terms of third order, such as $k_{x/y} k_{x/y}  k_z \sigma_{x/y/z}$, can impact the behaviour of other components for the optical conductivity but they do not contribute to the $zz$-component for which the leading order will always be $\sim \omega^3$.

The same argument holds for higher-order models where terms such as $k_z^5 \sigma_z$ and higher-order combinations of different $k$ components can be included. These contributions will only qualitatively change the behaviour of in-plane optical conductivity components. The leading order for the $zz$-component will remain the same, the prefactor can, however, change, especially if generic pre\-factors are included. In particular, this means that this statement also holds for tight-binding models with single topological nodal planes.

This discussion validates the relevance of the model studied in this paper for the characteristic low-frequency optical conductivity behaviour of topological nodal planes.

\section{Cubic frequency dependence in the generic case of arbitrary dispersions}\label{app:dispersion}

As described in the previous section, the non-symmorphic symmetry enforcing the nodal plane at \mbox{$k_z=0$} restricts the terms in the Hamiltonian to be odd in $k_z$.
One finds that the lowest order in the $zz$ component of the optical conductivity is $\sim \omega^3$.
In the following, we show that this result is still valid if one includes an arbitrary (symmetry-allowed) dispersion into the nodal plane low-energy expansion of Eq.\,\eqref{Eq:LE_NP}, i.e., $\mathcal{H}_\text{disp}(\mathbf{k}) = \mathcal{H}(\mathbf{k}) + d_0(\mathbf{k}) \mathbb{1}$ with some function $d_0(\mathbf{k})$.
As introduced in Eq.~\eqref{eq:linoptcond} we consider
\begin{align}
    \text{Re}[\sigma^{ab}] =  \frac{\pi \omega e^2}{4 \hbar (2 \pi)^3} \int \text{d}^3 k \delta \left(\omega - \frac{\Delta E}{\hbar}\right)  \Delta f\, \partial_a \Vec{n} \cdot \partial_b \Vec{n}\, .
\end{align}
The extra term $d_0(\Vec{k})\mathbb{1}$ neither influences $\Vec{n}$ nor $\Delta E$. It only takes effect on $\Delta f$, which is equal to 1 in the main text where for a finite gap the Fermi energy lies between the two bands. Further, note that the evaluation of the $\delta (\omega - \frac{\Delta E}{\hbar})$ term results in the integral being reduced to a surface integration. Solving $\omega - \frac{\Delta E}{\hbar}=0$ for $k_z$, we find that due to the presence of the symmetry-enforced nodal plane $k_{z,0}=a(k_x,k_y) \omega + \mathcal{O}(\omega^2)$ with some function $a(k_x,k_y)$, as long as the nodal plane is the only degeneracy of Eq.\,\eqref{Eq:LE_NP}. Then similarly, we get the low frequency expansion $\partial_z \Vec{n} \cdot \partial_z \Vec{n}=b(k_x,k_y)\omega^2 + \mathcal{O}(\omega^3)$ by plugging in the previous $k_z$ expansion. With $d_0(\mathbf{k})=0$ (the case from the main text) and these considerations, Eq.\,\eqref{eq:linoptcond} for $a=b=z$ becomes
\begin{align}
    \text{Re}[\sigma^{zz}] =  C \omega^3 \int \text{d}k_x\text{d}k_y h(k_x,k_y) +\mathcal{O}(\omega^4)\;,
\end{align}
with a constant $C$ and $h(k_x,k_y)= b(k_x,k_y) / |\partial_{k_z} \Delta E(\Vec{k})|_{k_z=k_{z,0}(\omega)}|$. Reintroducing a nonzero $d_0(\mathbf{k})$ implies the introduction of $\Delta f$
\begin{align}
    &\text{Re}[\sigma^{zz}] = C \omega^3 \nonumber\\ &\times\int \text{d}k_x\text{d}k_y h(k_x,k_y) \Delta f (k_x,k_y,k_z = k_{z,0}(\omega)) + \mathcal{O}(\omega^4)\;.
\end{align}
$\Delta f$ switches then between cases where transitions are allowed, $\Delta f = 1$, and those points where it is not, $\Delta f = 0$.  This is equivalent to reducing the integration area to an area $\Omega(\omega)$ in $k_x,k_y$ where $\Delta f = 1$, such that
\begin{align}
    \text{Re}[\sigma^{zz}] =  C \omega^3 \int_{\Omega(\omega)} \text{d}k_x\text{d}k_y h(k_x,k_y) +\mathcal{O}(\omega^4)\;.
\end{align}
Now, one needs to show, that the expansion of $\int_{\Omega(\omega)} \text{d}k_x\text{d}k_y h(k_x,k_y)$ is not $\omega^{-n}$ with $n>0$ in leading order, since this is the only way to reduce the order of the $\omega^3$-behavior. This is equivalent to saying that $\int_{\Omega(\omega)} \text{d}k_x\text{d}k_y h(k_x,k_y)$ remains finite for any $\omega$. This is indeed the case, since $h(k_x,k_y)\geq0$ and $\Omega(\omega)$ is a subset of $\mathbb{R}^2$, such that the $\Omega(\omega)$ integral is upper bounded by the $d_0(\Vec{k})=0$ case. Therefore, for any dispersion $d_0(\Vec{k})$, the leading order optical conductivity behavior in $\omega$ remains $\omega^3$.

\section{Analytical expression for the linear optical conductivity of the low-energy model}
\label{app:analytcond}
The full analytical result for the linear interband optical conductivity component perpendicular to the nodal plane $\sigma^{zz}$ for the case of $a\geq 0$ is given by
\begin{widetext}
    \begin{align}
        \sigma^{zz}_{a\geq0} =& \frac{\pi }{(2 \pi )^3} \frac{1}{9} \sqrt{\frac{2}{3}} \pi  \left(\frac{\left(f(\omega)-2 a \varepsilon ^2\right)^2}{\varepsilon ^2 f(\omega)}\right)^{3/2} - \frac{\pi }{(2 \pi )^3}\frac{2 \sqrt{\frac{2}{3}} \pi  \varepsilon ^2}{2835 \omega ^2} \left(\frac{\left(f(\omega)-2 a \varepsilon ^2\right)^2}{\varepsilon ^2 f(\omega)}\right)^{5/2}\nonumber \\
        &\hspace{2cm}\times\left( 63 a^2+\frac{15 a \left(f(\omega)-2 a \varepsilon ^2\right)^2}{\varepsilon ^2 f(\omega)}+
        \frac{35 \left(f(\omega)-2 a \varepsilon ^2\right)^4}{36 \varepsilon ^4 f(\omega)^{2}} \right)\; ,
    \end{align}
\end{widetext}
where $f(\omega) = \sqrt[3]{8 a^3 \varepsilon ^6+3 \left(\sqrt{48 a^3 \omega ^2 \varepsilon ^{10}+81 \omega ^4 \varepsilon ^8}+9 \omega ^2 \varepsilon ^4\right)}$.

\section{Tight-binding model of a topological nodal plane}
\label{app:tbmodel}
We consider the two-band tight-binding model of a topological nodal plane introduced in \cite{Xiao2020chargednodalsurface}, which can be written as
\begin{align}\label{eq_tb_Xiao}
    \hspace{1.cm}\mathcal{H}_\mathrm{tb} = d_0(\vb{k}) \mathbb{1} + \vb{d}(\vb{k}) \cdot \bm{\sigma}
\end{align}
with
\begin{align*}
    d_0(\vb{k}) &=  \left(2 \cos \left(\frac{k_x}{2}\right) \cos \left(\frac{\sqrt{3} k_y}{2}\right)+\cos (k_x)\right) \\
    &\hspace{2cm} \times 2 t_\mathrm{c} \cos (k_z) - \cos(k_z) + \mu \\
    d_x(\vb{k}) &=  \left(2 \cos \left(\frac{k_x}{2}\right) \cos \left(\frac{k_y}{2 \sqrt{3}}\right)+\cos \left(\frac{k_y}{\sqrt{3}}\right)\right)   \\
    &\hspace{2cm}\times 2 t_0 \cos \left(\frac{k_z}{2}\right)
\end{align*}
\begin{align*}
    d_y(\vb{k}) &=  -\left(2 \cos \left(\frac{k_x}{2}\right) \sin \left(\frac{k_y}{2 \sqrt{3}}\right)-\sin \left(\frac{k_y}{\sqrt{3}}\right)\right)   \\
    &\hspace{2cm} \times 2 t_0 \cos \left(\frac{k_z}{2}\right)  \\
    d_z(\vb{k}) &=  \left(\cos \left(\frac{\sqrt{3} k_y}{2}\right)-\cos \left(\frac{k_x}{2}\right)\right) \\
    &\hspace{2cm} \times 4 t_\mathrm{c} \sin \left(\frac{k_x}{2}\right) \sin (k_z)  \; ,
\end{align*}
where $\bm{\sigma}$ is the vector of Pauli matrices. We added the $-\cos(k_z)$ term and the chemical potential $\mu$ in $d_0$ to move the Weyl points far below the Fermi energy such that their contribution to the optical conductivity is suppressed.  For our calculation we use the parameters $t_0 =0.4$ and $t_\mathrm{c} = 0.1$ and we set the chemical potential to $\mu = -1.24$. The resulting band structure is shown in Fig.~\ref{fig:tblinearcond_zz}(a).

By expanding the Hamiltonian (\ref{eq_tb_Xiao}) around the high-symmetry point H $= (\frac{4\pi}{3}, 0, \pi)^\mathrm{T}$ one obtains the low-energy model
\begin{align}
    \mathcal{H}_\mathrm{tb,LE} = \frac{\sqrt{3}}{2}t_0 k_x k_z \sigma_x &- \frac{\sqrt{3}}{2}t_0 k_y k_z \sigma_y + \nonumber\\
    &+\left(\frac{\sqrt{3}}{2}t_c k_z^3 - b\; 3\sqrt{3} t_c k_z\right)\sigma_z ,
\end{align}
which allows us to understand the dependence of the analytical solution on the parameters of the tight-binding model. Note, that for this we neglected further terms cubic in $k$, which are expected to only give small corrections in the prefactors. Furthermore, the energy offset caused by the dispersion and the chemical potential is neglected. This low-energy model corresponds to the $a=-1$ phase discussed for the model in Eq.~(\ref{Eq:LE_NP}), which means that two additional Weyl points are present. In order to remove their contribution to the optical response, we change the phase to correspond to the $a=+1$ case by introducing the parameter $b$ and setting $b=-1$. With this, the analytical behaviour shown in Fig.~\ref{fig:tblinearcond_zz}(b) is obtained.

\section{Trivial nodal plane}
\label{app:trivNP}
The simplest low-energy model for a topologically trivial nodal plane, for which the quantum metric does not vanish, is given by
\begin{align}
    \mathcal{H}_\mathrm{triv} = k_z \sigma_x + k_z \sigma_y + (k_z + k_z^3)\sigma_z \; .
\end{align}
The only non-zero component of the QGT in the model is the quantum metric component perpendicular to the nodal plane given by
\begin{equation}
    g^{zz}_\pm = \frac{2 k_z^2}{(3+2k_z^2 + k_z^4)^2}\; .
\end{equation}
As for the topological model, we calculate the $zz$-component of the linear optical conductivity. Since there is no dispersion in $k_{x/y}$-direction in this model, the integral over these $k$-components diverges, which we avoid by normalizing the integral by the diverging $k_x$-$k_y$-volume. The resulting optical conductivity is given by

\begin{widetext}
    \begin{align}
        \sigma^{zz} = \frac{\pi}{(2\pi)^3 }\left[432 \omega ^2 \left(f(\omega)-\frac{20}{f(\omega)}-4\right)\right]\bigg/
        \left[\left(4 f(\omega)+\left(f(\omega)\right)^{2}
        -\frac{80}{f(\omega)}+\frac{400}{\left(f(\omega)\right)^{2}}+36\right)^2\right.\nonumber\\
        \times\left.\left(\frac{2}{3} \sqrt{\frac{2}{3}} \left(f(\omega)-\frac{20}{f(\omega)}-4\right)^{3/2}+
         \frac{\left(f(\omega)-\frac{20}{f(\omega)}-4\right)^{5/2}}{6 \sqrt{6}}+\sqrt{6} \sqrt{f(\omega)-\frac{20}{f(\omega)}-4}\right)\right] \; ,
    \end{align}
\end{widetext}
where $f(\omega) = \sqrt[3]{27 \omega ^2+3 \sqrt{81 \omega ^4+912 \omega ^2+3456}+152}$.
By expanding the expression around $\omega = 0$, we obtain again a leading cubic order
\begin{align}
    \sigma_{zz} = \frac{\pi}{(2\pi)^3 }\frac{\omega ^3}{54 \sqrt{3}} -\frac{\pi }{(2\pi)^3 }\frac{\omega ^5}{216 \sqrt{3}}+ \mathcal{O}\left(\omega ^7\right)\; ,
\end{align}
which shows that this low-frequency behaviour does not depend on the topological charge but is a characteristic feature of nodal planes. In particular, every Berry curvature component in this model vanishes, demonstrating again the important role of the quantum metric for the investigation of response properties of materials.

\section{Suitable space groups and materials}
\label{app:Materials}
In order to study the low-frequency optical response of a nodal plane material experimentally, one can select systems with certain properties, which are expected to lead to a more reliable determination of the power-law behaviour.
In general, there are two requirements
that a nodal plane material must satisfy in order to show a clear $\omega^3$
power law in its optical conductivity:

The first requirement is that
the Drude peak, originating from intraband transitions of the nodal-plane bands (and possibly other bands near the Fermi energy),
should be narrow in frequency.
In general, for a narrow Drude peak, clean systems are required, which should be measured at low temperature, while making sure that the crystal structure is stable at low temperatures and no magnetic order develops. Furthermore, for the Drude peak to be small, no other band features should be located at the Fermi level apart from the nodal plane and the density of states should be low.
We would like to note, however,
that in principle it is possible to subtract the Drude contribution
from the optical conductivity, to reveal the interband response,
as recently demonstrated for the multi-fold fermion compound
 RhSi \cite{Maulana2020RhSi}.

The second requirement is a broad frequency range for which the cubic power-law behaviour can be measured. As for the first point, this is expected to be optimal in systems with no other bands (and most importantly topological crossings) at the Fermi level. In particular, the energy range in which the two bands of the nodal plane disperse away from it without intersecting with each other or other bands 
should be as large as possible. Since in the case of a topological nodal plane the two bands need to form Weyl points in order to compensate the topological charge, trivial nodal plane systems are expected to be better suited for this. All of the discussed requirements can ensure a large frequency range, in which the nodal plane signature is expected to be visible.

With these criteria, we find that relevant space groups (SGs) should have only one two-fold screw rotation (for a single nodal plane) and mirror and/or inversion symmetries (for a trivial charge). We list all SGs which can host single nodal planes in Tab.~\ref{tab:SGs_materials}, where we also indicate, whether spin-orbit-coupling (SOC) can gap out the nodal plane and whether a topological charge is allowed for the specific SG. With this, one can perform a search for previously synthesized materials, which fit our criteria to a reasonable degree and which we propose to be suitable for the optical conductivity measurement. In particular, we suggest that Sb$_3$Sr$_5$ (SG 193) \cite{Martinez73SbSr} and $X$Mo$_3$S$_3$ ($X$ = Rb, K, Cs) (SG 176) \cite{Wu2018,Huster1983333} are the best candidates for this. The latter material group is particularly promising, since these materials have an almost flat trivial nodal plane as the only feature at the Fermi level. We further propose HfI$_3$ (SG 193) \cite{Dahl64HfI} and K$_4$P$_3$ (SG 63) \cite{Schnering89KP} as suitable materials.

\begin{table}[h!]
    \centering
    \begin{tabular}{r|c|c|c|c}
        \hline\hline
        \multicolumn{2}{c|}{SG}   & NP position & NP w. SOC & top. NP  \\
        \hline
        4 & $P 1 \, 2_1 \, 1$ & $k_y = \pi$ & \checkmark & allowed  \\
        11 & $P 1 \, 2_1/m \, 1$ & $k_y = \pi$ & -- & no  \\
        14 & $P 1 \, 2_1/c \, 1$ & $k_y = \pi$ & -- & no  \\
        17 & $P 2 2 2_1$ & $k_z = \pi$ & \checkmark & allowed  \\
        20 & $C 2 2 2_1$ & $k_z = \pi$ & \checkmark & allowed  \\
        26 & $P m c 2_1$ & $k_z = \pi$ & \checkmark & no  \\
        29 & $P c a 2_1$ & $k_z = \pi$ & \checkmark & no  \\
        31 & $P m n 2_1$ & $k_z = \pi$ & \checkmark & no  \\
        33 & $P n a 2_1$ & $k_z = \pi$ & \checkmark & no  \\
        36 & $C m c 2_1$ & $k_z = \pi$ & \checkmark & no  \\
        51 & $P 2_1/m \, 2/m \, 2/a$ & $k_z = \pi$  & -- & no  \\
        52 & $P 2/n \, 2_1/n \, 2/a$ & $k_z = \pi$ & -- & no  \\
        53 & $P 2/m \, 2/n \, 2_1/a$ & $k_z = \pi$ & -- & no  \\
        54 & $P 2_1/c \, 2/c \, 2/a$ & $k_z = \pi$ & -- & no  \\
        63 & $C 2/m \, 2/c \, 2_1/m$ & $k_z = \pi$ & -- & no  \\
        64 & $C 2/m \, 2/c \, 2_1/e$ & $k_z = \pi$ & -- & no  \\
        76 & $P 4_1$ & $k_z = \pi$ & \checkmark & allowed  \\
        78 & $P 4_3$ & $k_z = \pi$ & \checkmark & allowed  \\
        91 & $P 4_1 2 2$ & $k_z = \pi$ & \checkmark & allowed  \\
        95 & $P 4_3 2 2$ & $k_z = \pi$ & \checkmark & allowed  \\
        169 & $P 6_1$ & $k_z = \pi$ & \checkmark & allowed  \\
        170 & $P 6_5$ & $k_z = \pi$ & \checkmark & allowed  \\
        173 & $P 6_3$ & $k_z = \pi$ & \checkmark & allowed  \\
        176 & $P 6_3/m$ & $k_z = \pi$ & -- & no \\
        178 & $P 6_122$ & $k_z = \pi$ & \checkmark & allowed  \\
        179 & $P 6_522$ & $k_z = \pi$ & \checkmark & allowed  \\
        182 & $P 6_322$ & $k_z = \pi$ & \checkmark & allowed  \\
        185 & $P 6_3 c m$ & $k_z = \pi$ & \checkmark & no  \\
        186 & $P 6_3 m c$ & $k_z = \pi$ & \checkmark & no  \\
        193 & $P 6_3/m \, 2/c \, 2/m$ & $k_z = \pi$ & -- & no \\
        194 & $P 6_3/m \, 2/m \, 2/c$ & $k_z = \pi$ & -- & no  \\
         \hline\hline
    \end{tabular}
    \caption{All space groups with single nodal planes. First and second column: Space group (SG) number and full symbol. Third column: Position of the nodal plane. Fourth column: Check mark denotes SGs, for which spin-orbit-coupling (SOC) splits the spin-degenerate nodal plane (without SOC) into two nodal planes, otherwise the nodal planes are gapped with SOC. Fifth column: In the absence of mirror symmetries, a nodal plane may carry a non-zero Chern number, denoted as ``allowed" topological nodal plane.}
    \label{tab:SGs_materials}
\end{table}

\bibliography{bibliography.bib}

\end{document}